\documentclass[final,5p, times, 12pt]{elsarticle}
\usepackage{amssymb}
\usepackage{amsthm}
\usepackage{xcolor}
\usepackage{physics}
\usepackage[colorlinks=true,linkcolor=black,citecolor=black]{hyperref}
\usepackage{ulem}
\usepackage{setspace}
\usepackage[hang,flushmargin]{footmisc}

\biboptions{sort&compress}
\usepackage{subfigure}
\DeclareUnicodeCharacter{2009}{\,} 
\journal{Carbon}

\makeatletter
\def\thefootnote{\fnsymbol{footnote}}
\makeatother
\usepackage{ragged2e}

\justifying
\begin{document}

\begin{frontmatter}

\title{\texorpdfstring{Enhanced magnetic field sensitivity of shallow NV$^-$ ensembles via high-temperature implantation}{Enhanced magnetic field sensitivity of shallow NV- ensembles via high-temperature implantation}}

\author[inst1]{Joa Al Yahya \footnotemark[1]\footnotemark[2]}
\author[inst1]{Anatole Bach \footnotemark[1]\footnotemark[2]}
\author[inst1]{Jayash Panigrahi}
\author[inst1]{Pauline Perrin}
\author[inst1]{Ionut Balasa}
\author[inst1]{Diana Serrano}
\author[inst1]{Alexey Tiranov}
\author[inst2]{Jocelyn Achard}
\author[inst1]{Alexandre Tallaire}
\author[inst1]{Philippe Goldner}

\affiliation[inst1]{organization={Chimie ParisTech, PSL University, CNRS, Institut de Recherche de Chimie Paris},
            addressline={75005 Paris}, 
            country={France}}
\affiliation[inst2]{organization={Sorbonne Paris Nord University, CNRS, Laboratoire des Sciences des Procédés et des Matériaux},
            addressline={93430 Villetaneuse}, 
            country={France}}

\begin{abstract}
Dense and shallow ensembles of negatively charged nitrogen-vacancy centers (NV$^-$) with good optical and spin properties play a key role in the performance enhancement of diamond-based quantum sensors. Ion implantation enables precise control of NV$^-$ depth and density. However, at high ion fluence, this method is limited by low NV$^-$ creation yields and sample amorphization. Additionally, shallow NV$^-$ spin properties deteriorate due to surface proximity. In this paper, we study N$_2^+$ ion implantation at energies between 10 and 15 keV with fluences as high as 1$\times$10$^{15}$ ions/cm$^2$   at temperatures of 20, 400 and 800°C  to investigate the influence of implantation temperature on lattice damage, NV$^-$ creation yield and NV$^-$ spin properties. Our results show that diamond maintains structural integrity at 800°C with fluences up to 1$\times$10$^{15}$ ions/cm$^2$ without amorphization. Furthermore, high-temperature implantation improves NV$^-$ creation yields up to five times without compromising  T$_2^*$, T$_2$ and T$_1$, making it a promising approach to enhance the magnetic field sensitivity of NV$^-$ ensembles.

\end{abstract}


\end{frontmatter}

\section{Introduction}
\makeatletter
\def\thefootnote{\fnsymbol{footnote}}
\makeatother
\footnotetext[1]{joa.morla-al-yahya@chimieparistech.psl.eu\newline 
     anatole.bach@chimieparistech.psl.eu}
\footnotetext[2]{These authors contributed equally to this work.}
Among the various point defects in diamond, the negatively charged nitrogen-vacancy  center (NV$^{-}$) stands out for its optical and spin properties at room temperature \cite{barry_sensitivity_2020}. 
The spin state of the NV$^{-}$ can be controlled and read out using off-resonant optical pulses.
This unique characteristic makes NV$^{-}$ highly suitable for a wide range of applications, including quantum computing \cite{abobeih_fault-tolerant_2022}, quantum communication, and quantum sensing of magnetic fields \cite{taylor_high-sensitivity_2008}, electric fields, and temperature \cite{kucsko_nanometre-scale_2013}. 

The sensitivity of diamond NV$^{-}$ based magnetometers depends strongly on the density, spatial distribution, and spin properties of these centers. Achieving high sensitivity requires the creation of shallow and dense NV$^{-}$ ensembles with good spin lifetime and coherence. 
Focused ion beam (FIB) implantation enables precise positioning of shallow NV$^{-}$ within the diamond. 
The depth of the NV$^{-}$ can be controlled by the ion beam energy, while their density depends on the ion beam fluence.

During the implantation process, the interaction of the nitrogen ions with the lattice generates vacancies and interstitial carbon atoms along their path. At low implantation fluences, the formation of NV$^{-}$ is limited by the amount of available vacancies. As the sample is annealed after implantation at temperatures above 600°C, vacancies become mobile and diffuse to substitutional nitrogen atoms to form NV$^{-}$. However, at high ion fluences, the large amount of vacancies leads to the formation of vacancy clusters and causes lattice damage. When the vacancy density exceeds $2.8\times10^{22}$ vacancies/cm$^{3}$, it leads to amorphization of the diamond structure \cite{fairchild_mechanism_2012,uzan-saguy_damage_1995}. 
This sets an upper limit on the achievable density of NV$^{-}$ \cite{pezzagna_creation_2010}.
In addition, the maximum density of shallow NV$^-$ ensembles is further constrained by the decrease in NV$^-$ creation yields for low-energy ion implantation. While 50\% of the nitrogen atoms implanted in the MeV energy range form NV$^-$, creation yields drop to 2\% for implantation at 10 keV. \cite{pezzagna_creation_2010} These challenges limit the achievable density of shallow NV$^{-}$.

Recent work has demonstrated a 3 to 10-fold creation yield enhancement for implanted single NV$^-$, using charge-assisted defect engineering. Diamond doping with phosphorus, oxygen, or sulfur atoms helps to stabilize the NV$^-$ charge while hindering the formation of other vacancy defects. However, this increase in creation yield drops with the nitrogen fluence, as the amount of implantation-induced vacancies surpasses the number of dopants \cite{luhmannCoulombdrivenSingleDefect2019}.

It was previously shown in our group that heating the diamond substrate during FIB implantation can mitigate the formation of vacancy clusters \cite{capelli_increased_2019,ngandeu_ngambou_hot_2024} and allow a higher density of NV$^{-}$ without compromising the structural integrity of the diamond lattice.
In this work, we extend these findings by systematically examining the impact of substrate heating on NV$^{-}$ creation yield and spin coherence across a broad range of implantation parameters.
We investigate implantation at 20, 400 and 800°C for various nitrogen ion energies and fluences and study both the optical and spin properties of the NV$^{-}$ ensembles. 
We show that implanting at 800°C prevents the sample amorphization and enhances the NV$^-$ creation yield up to 5 times while preserving the NV$^{-}$ spin properties - resulting in enhanced sensitivity to both AC and DC magnetic fields. Heating the substrate at 400°C during implantation also enhances the creation yield, although the vacancies are not mobile at this temperature.
This enhancement is attributed to the release of carbon interstitials trapped by substitutional nitrogen atoms.

The paper is organized as follows. 
Section 2 presents the sample growth method and the experimental techniques used to characterize the sample. 
In Section 3.1, we examine the effects of substrate heating during FIB implantation on structural damage and creation yield. 
Finally, section 3.2 discusses the spin properties of NV$^{-}$ as a function of implantation temperature.

\section{Methods}

\subsection{Sample growth}

A 20-µm thick ultra-high purity diamond layer was grown by microwave plasma-assisted chemical vapor deposition (MPACVD) on a type Ib High-Pressure High-Temperature (HPHT) diamond substrate using a home-made reactor. A MW power of about 3 kW and a pressure of 200 mbar with the addition of 4\% CH$_4$ to H$_2$ was used. More details on the growth conditions can be found in a previous paper \cite{achardHighQualityMPACVD2007}.

NV$^-$ were implanted using a focused ion beam system optimised for ion implantation (QuiiNN, \textit{Orsay Physics}). The system is equipped with a plasma source (iVeloce, \textit{Orsay Physics}) and a Wien filter allowing precise mass separation at energies ranging from 5 to 30 keV. A $^{15}$N$_2^+$ ion beam was selected and scanned across a designated area on the surface. The implantation dose was calculated from the current measured using an in-situ Faraday cup. The sample was mounted onto a heated stage (FurnaSEM 1000, \textit{NewTec}) so that implantation could be carried out at temperatures up to 950°C.

Fluences of 10$^{14}$, 2x10$^{14}$, 4x10$^{14}$, and 1x10$^{15}$ N/cm$^2$ were used at energies of 10, 12.5, and 15 keV (per nitrogen atom). The corresponding NV- average depths, estimated using Stopping and Range of Ions in Matter (SRIM) simulations \cite{ziegler_srim_2010} (see Appendix A), were 15, 18, and 22 nm, respectively. During implantation the temperature of the stage was set to either 20, 400 or 800°C. Annealing was then performed ex-situ in a lamp furnace (AS-ONE, \textit{Annealsys}) at 850 °C under a high vacuum of about 10$^{-5}$ mbar for 1 h.

\subsection{Experimental methods}

Sample surface images were captured using an optical microscope (BX43, \textit{Olympus}) in transmission mode.

Raman spectra under 633 nm optical excitation were performed using a Raman spectrometer (\textit{Renishaw InVia}) with a 50X objective and 1800 grooves/mm grating. \color{black}

Photoluminescence (PL) and spin experiments were performed using a confocal microscopy setup. NV$^-$ were optically excited using a diode-pumped solid-state green laser (DJ520-40, \textit{Thorlabs}), which can be modulated using two acousto-optic modulators (\textit{AA Opto-Electronic}) to generate pulses.
The excitation beam was focused onto the sample using a 50X objective with a 0.95 numerical aperture (NA). 
The sample was mounted on a piezoelectric positioner (P-611.3 NanoCube, \textit{Physik Instrumente}) which allowed NV$^-$ PL mapping.
The NV$^-$ PL was spectrally filtered by a 650-800 nm band-pass filter and collected through a confocal setup and subsequently focused onto a single photon avalanche photodiode (COUNT10-C, \textit{Laser Components}) interfaced with a time-to-digital converter (quTAG, \textit{qutools}).

Microwave pulses were generated  (SG38X, \textit{Stanford Research Systems}), amplified (ZHL-15W-422-S+, \textit{Mini-Circuits}),  and directed to a custom-built 100-µm diameter wire antenna placed near the sample surface. 
A permanent magnet produced a field strength of 1 mT at the sample surface, in order to lift the degeneracy of the NV$^-$ \(\ket{m_s = \pm1}\)  spin states, enabling addressing individual spin transitions. 
Linearly polarized excitation was employed to enhance the contrast of the spin transition of interest, as detailed in \ref{appendix:pulselengths}. 
All pulse sequences required for optical and microwave excitation were programmed and executed using a pulse and delay generator (PCIe Board SP46, \textit{Spincore}).

 \section{\texorpdfstring{Results and Discussion}{Optical properties of the NV- ensemble}}

\subsection{Optical properties of the NV$^{-}$ ensemble}
\subsubsection{Diamond amorphization}
\begin{figure*}
    \centering
    \includegraphics[width=\textwidth]{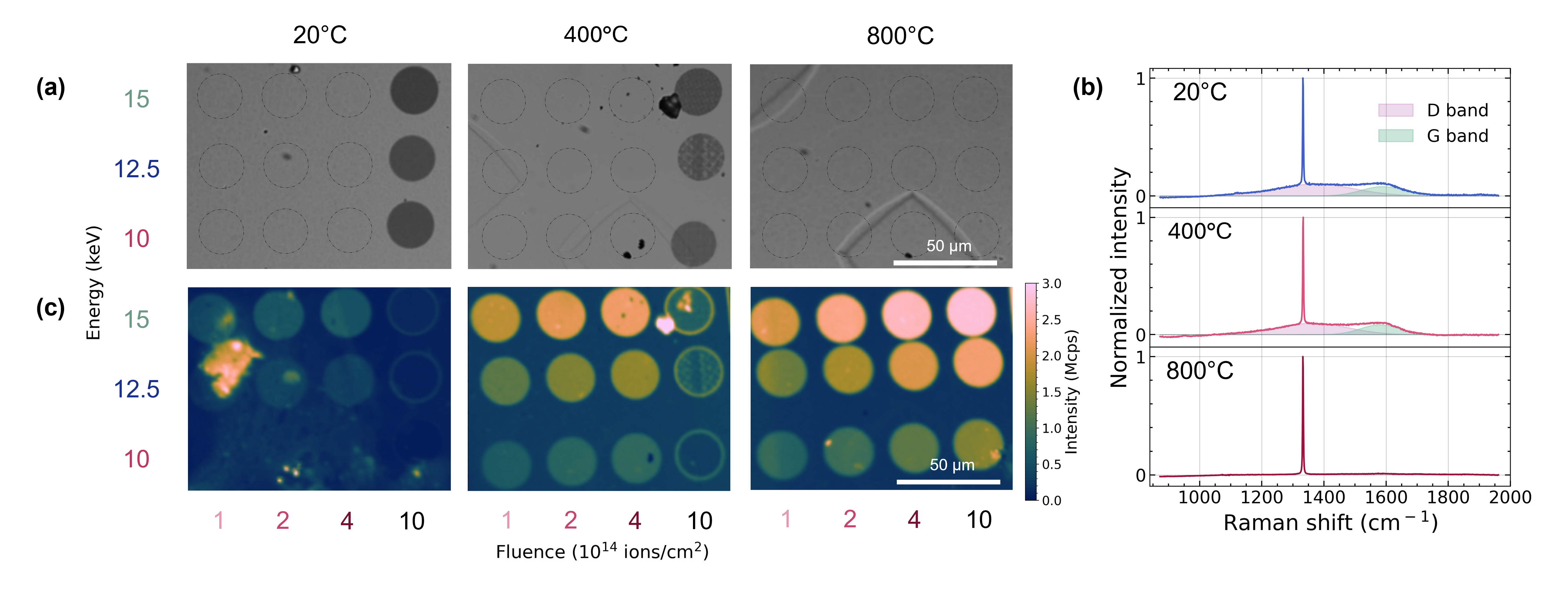}
    \caption{(a) Optical microscopy images (transmission mode) of the diamond surface after N$_2^+$ implantation at 20, 400 and 800°C with varying ion fluences and energies. The darkened spots correspond to amorphized regions. The dotted circles indicate the regions where the ions were implanted. (b) Raman spectra in areas implanted at 20, 400 and 800 °C with a fluence of 1$\times$10$^{15}$ ions/cm$^2$ and an energy of 15 keV (22 nm). The contributions of the graphite G  (green) and disorder D (purple) bands are fitted by Gaussian functions. (c) Photoluminescence of the NV$^-$ implanted at 20, 400 and 800°C with varying ion fluences and energies.}
    \label{fused}
\end{figure*}

Optical images of the sample surface are presented in Figure \ref{fused}(a).
For nitrogen implantation  at 20 and 400°C, areas implanted at 1$\times$10$^{15}$ ions/cm$^2$ exhibit darkened spots indicating amorphization of the diamond. In contrast, no visible sign of amorphization is observed in the region where implantation was carried out at 800°C.

To further investigate diamond damage, Raman spectra were taken after implantation at 1$\times$10$^{15}$ ions/cm$^2$ with an energy of 15 keV at 20, 400 and 800°C, and are presented in Figure \ref{fused}(b). 
The 800°C spectrum only exhibits the characteristic diamond Raman peak at \(1332\ \text{cm}^{-1}\), while those for implantation at 20 and 400°C show additional features corresponding to the graphite (G) and the disorder (D) bands, both indicative of amorphous carbon \cite{dychalska_study_2015}. 

These observations confirm that high-fluence implantation induces amorphization of the diamond lattice when performed at 20 and 400°C.
\color{black}
This does not occur when the sample is heated at 800°C during implantation. At 800°C, single vacancies are mobile \cite{allers_annealing_1998}, which favors their diffusion out of the damaged areas, as well as vacancy-interstitial recombinations. This prevents the formation of multi-vacancy clusters responsible for lattice damaging and amorphization. 
Therefore, implanting at temperatures at which vacancy mobility is activated helps preserve the structural integrity of the diamond lattice, even at high fluences.

\subsubsection{Creation yield}

The PL maps of the NV$^-$ implanted at 20, 400, and 800°C are presented in Figure \ref{fused}(c). 
For any given implantation energy and fluence, the NV$^-$ spots exhibit an increase in PL when the substrate is heated during implantation. Moreover, there is an absence of NV$^-$ PL for implantation with a fluence of $1\times10^{15}$ ions/cm$^2$ at 20 and 400°C. This confirms the amorphization of the diamond lattice during high-fluence implantation at temperatures where vacancies are immobile.

We estimate that the PL increase displayed in Figure \ref{fused}(c) indicates that more NV$^-$ are created when the substrate's temperature is increased during implantation, assuming that PL is proportional to the amount of NV$^-$. To quantify the improved creation yield $Y$, we used the following formula \cite{pezzagna_creation_2010} :  
\begin{equation}
Y = \frac{I}{I_{NV} \times F }
\end{equation}
where I is the PL intensity per unit area of the NV$^-$ ensemble,  I$_{NV}$ the PL intensity of a single NV$^-$ which was determined using a reference sample (see \ref{Appendix E}), and F is the ion beam fluence. 

\begin{figure}[ht]
    \centering
    \includegraphics[width=0.7\columnwidth]{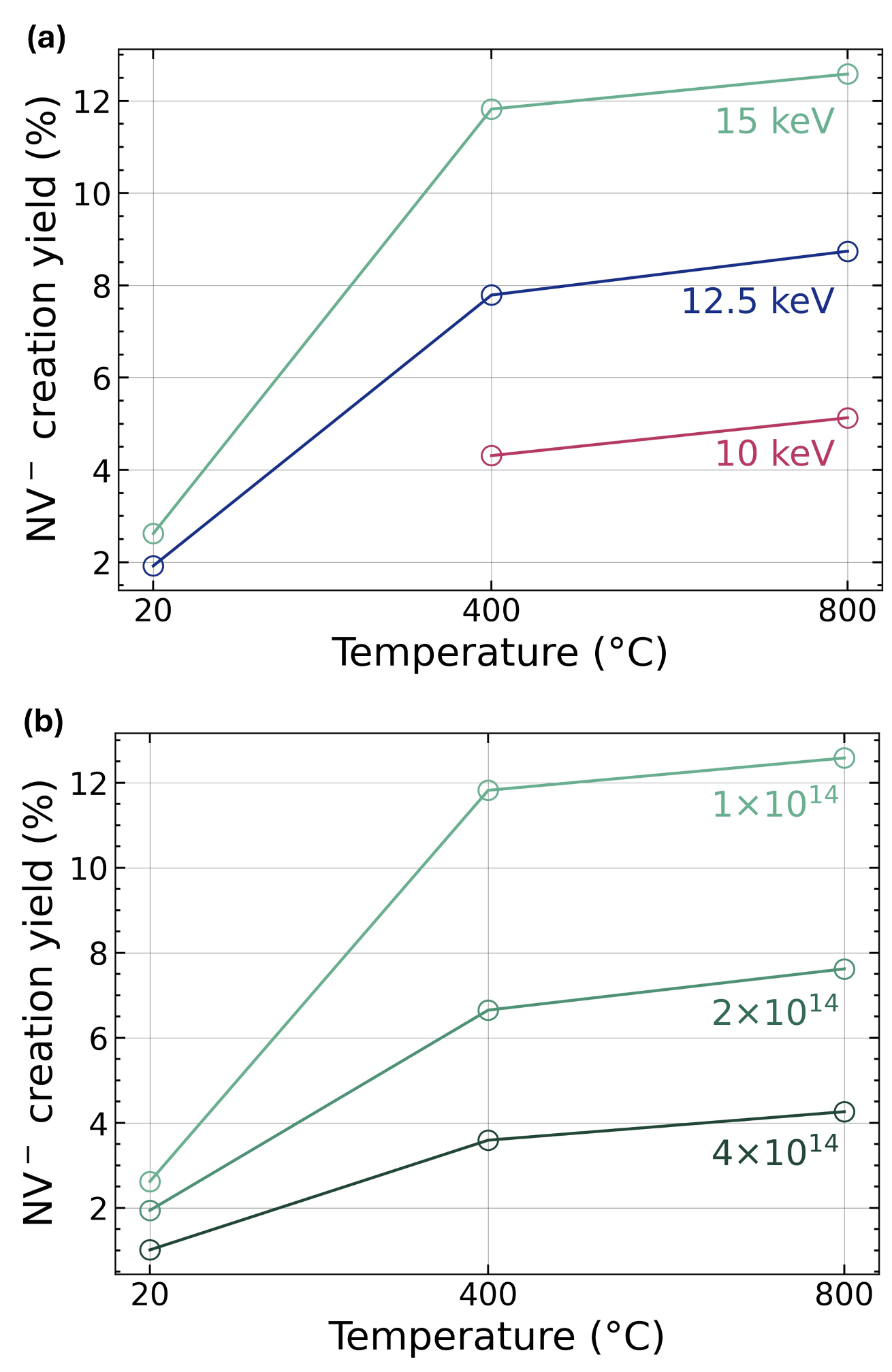}
    \caption{NV$^-$ creation yield as a function of temperature for a) varying implantation energies with a fluence of 1$\times$10$^{14}$ ions/cm$^2$ and b) varying fluences with an implantation energy of 15 keV (22 nm).}
    \label{fig:PLyield}
\end{figure}

The variation of $Y$ as a function of the implantation temperature is represented for various energies at a 1$\times$10$^{14}$ ions/cm$^2$ fluence in Figure \ref{fig:PLyield}(a) and for various ion fluences for a 15 keV ion energy in Figure \ref{fig:PLyield}(b).
For all implantation fluences and energies, we observe a significant increase in the NV$^-$ creation yield from 20 to 400°C and a smaller increase between 400 and 800°C. Around 3\% of the nitrogen implanted at 20°C with an ion beam energy of 15 keV and a fluence of 1$\times$10$^{14}$ ions/cm$^2$ recombine to form a NV$^-$, which is in good agreement with creation yields found by Pezzagna et al \cite{pezzagna_creation_2010}. For the same implantation conditions in a 800°C-heated diamond, the creation yield Y is close to $\sim$13\%.
Implantation at 800°C mitigates the formation of vacancy clusters, resulting in a higher amount of mobile single vacancies available to diffuse to substitutional nitrogen, both during implantation and annealing.

Although vacancies are not mobile below 600°C, the increase in the NV$^-$ creation yield also occurs for nitrogen implantation at 400°C.
Vacancy mobility is therefore not the only mechanism leading to NV$^-$ creation yield increase.
A potential explanation resides in carbon interstitials and their interactions with substitutional nitrogen. 
Carbon interstitials get trapped by substitutional nitrogen atoms at low temperatures, and can only get released through annealing at 400 °C \cite{iakoubovskii_annealing_2003}. This makes nitrogen recombination with a vacancy less likely below 400°C than above. Implantation at 400°C would thus result in higher creation yields than 20°C implantation.

While Figure \ref{fig:PLyield}(a) shows the expected increase in creation yield with rising implantation energy, Figure \ref{fig:PLyield}(b) reveals a decrease in creation yield with increasing fluence. This was previously observed by Pezzagna et al. \cite{pezzagna_creation_2010}, suggesting that the yield could be further enhanced at high implantation temperature at lower fluences.

\subsection{\texorpdfstring{Spin properties of the NV$^{-}$ ensemble}{Spin properties of the NV- ensemble}}
\label{section2}

\subsubsection{Spin dephasing time}

To estimate NV$^-$ spin dephasing times $T_{2}^*$, we recorded continuous-wave ODMR spectra under a small magnetic field with a microwave excitation power of 4 dBm, as described in \ref{appendix:pulselengths}. Figure \ref{fig:SpinNVcombined_green}(a) show the results for NV$^-$ implanted at an energy of 15 keV, a fluence of 1$\times$10$^{14}$ ions/cm$^2$, and for the three implantation temperatures. 
The microwave power was chosen to correspond to the point just before the linewidth starts to increase.
\begin{figure*}[ht]
    \centering
    \includegraphics[width=\textwidth]{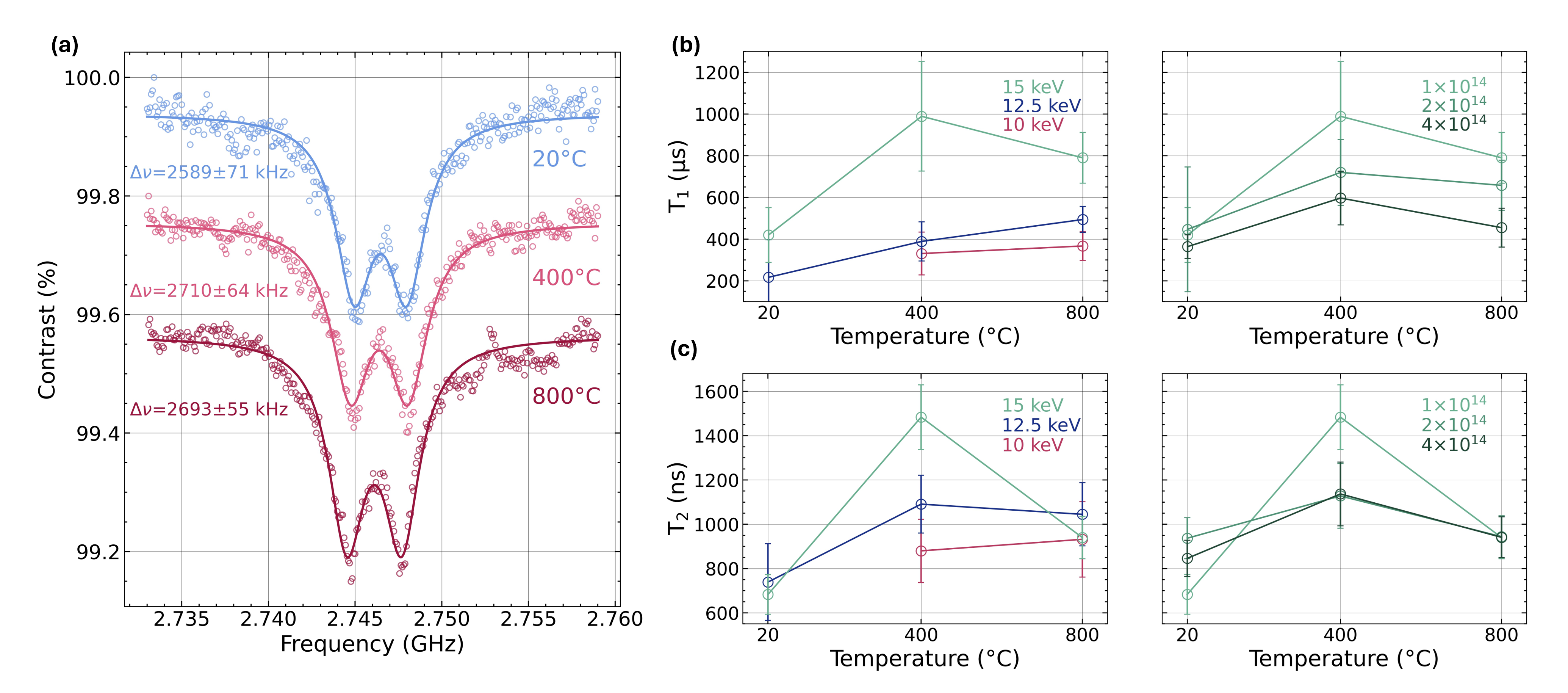}
    \caption{(a) ODMR spectra of the NV$^{-}$ implanted at 20, 400 and 800°C with an energy of 15 keV (22 nm) and a fluence of 10$^{14}$ ions/cm$^2$. (b) Variation of  T$_1$ with temperature for varying implantation energies with a fluence of 1$\times$10$^{14}$ ions/cm$^2$ (left panel) and for varying fluences with an implantation energy of 15 keV (right panel). (c) Variation of the spin coherence time T$_2$ with temperature for varying implantation energies with a fluence of 1$\times$10$^{14}$ ions/cm$^2$ (left panel) and for varying fluences with an implantation energy of 15 keV (right panel).}    
    \label{fig:SpinNVcombined_green}
\end{figure*}
As expected, the ODMR spectra exhibit the characteristic energy doublet associated with $^{15}$N nuclear spins ($I=1/2$). The linewidths of the resonance peaks are 2.6 MHz at 20°C, 2.7 MHz at 400°C, and 2.7 MHz at 800°C, with respective contrasts of 0.28\%, 0.27\%, and 0.3\%.

Importantly, while strongly increasing the amount of NV$^-$, implantation at high temperature does not result in a broadening of the ODMR linewidth or a decrease in contrast. This indicates an improvement of the magnetic sensitivity with implantation temperature, up to a factor of 4 to 5 at 800 ºC.
As demonstrated in the previous section, when implanting at 400°C and 800°C, more NV$^-$ are formed and thus less substitutional nitrogen is present in the sample. As both these paramagnetic defects participate in the spin bath, it is expected that $T_{2}^*$ does not decrease.

\subsubsection{Spin population lifetime}

The spin relaxation time \(T_1\) of the NV$^-$ ensembles was measured using the pulse sequence described in \ref{appendix:pulselengths}. 
The variation of T$_1$ as a function of the implantation temperature is represented in Figure \ref{fig:SpinNVcombined_green}(b) for various energies when the ion fluence is 1$\times$10$^{14}$ ions/cm$^2$  (left panel) and for various ion fluences when the ions energy is 15 keV (right panel).
For all implantation fluences and energies, T$_1$ increases from 20 to 400°C and is approximately the same at 400 and 800°C.

In low-density and deep NV$^-$ ensembles, $T_1$ is dominated by spin-lattice relaxation which is driven by phonon interactions \cite{norambuena_spin-lattice_2018}. 
It  includes processes such as two-phonon Raman scattering and the Orbach mechanism, which typically exhibit strong temperature dependence due to their reliance on thermal phonon populations. 
However, in high-density shallow NV$^-$ ensembles, such as those studied here, other relaxation mechanisms become important.
First, the high density of NV$^-$ can lead to cross-relaxation within the spin ensemble and, second, shallow positioning of NV$^-$ near the surface increases the effect of electric and magnetic surface noise. 
The former is characterized by fast depolarization of some spins within the ensemble caused by nearby defects, allowing spin relaxation via flip-flop processes \cite{choi_depolarization_2017,pellet-mary_relaxation_2023}. 
The latter effect arises from the surface roughness, which introduces charge traps. These traps, commonly located at the sample surface, can capture electrons from NV$^-$, playing a crucial role in spin decay mechanisms \cite{myers_double-quantum_2017}. 
As illustrated in the left panel of Figure \ref{fig:SpinNVcombined_green}(b), surface noise manifests as a systematic reduction in the lifetime of shallow NV$^-$ ensembles relative to those located deeper within the diamond. 

We observe that despite an increased concentration of centers at elevated implantation temperatures, T$_1$ doesn't decrease, which suggests  that any enhancement in cross-relaxation is effectively counterbalanced by high-temperature implantation, which reduces the impurities in the diamond lattice and therefore the number of rapidly depolarizing spin among the ensemble.

\subsubsection{Coherence time}

The coherence time \(T_2\) of the NV$^-$ ensemble was measured using the Hahn echo pulse sequence, as described in \ref{appendix:pulselengths}.
The variation of T$_2$ as a function of the implantation temperature is represented in Figure \ref{fig:SpinNVcombined_green}(b) for various energies when the ion fluence is 1$\times$10$^{14}$ ions/cm$^2$  (left panel) and for various ion fluences when the ions energy is 15 keV (right panel).
For all implantation fluences and energies, the coherence time increases from 20 to 400°C and is roughly the same at 400 and 800°C. 

\(T_2\) is predominantly degraded by the presence of paramagnetic defects within the diamond lattice, with the \(P1\) center, i.e. substitutional nitrogen, being the most significant contributor \cite{bauch_decoherence_2020}. High densities of \(P1\) centers lead to increased dipolar interactions, resulting in a substantial reduction in \(T_2\). 
For shallow NV$^-$ ensembles, the proximity of the surface introduces additional sources of magnetic noise. 

Implanting at higher temperatures increases the NV$^-$ creation yield and therefore diminishes the concentration of P1 centers, resulting in a higher coherence time. The annealing of other paramagnetic defects mentioned in the previous subsection might also contribute to the increase in coherence time.

\section{Conclusions}

In this work, we demonstrated that high-temperature focused ion beam (FIB) implantation significantly improves the creation yield and spin properties of dense and shallow NV$^-$ ensembles in diamond. By varying the implantation temperature, we showed that heating the substrate to 800°C prevents the amorphization of the diamond lattice, even at high ion fluences, enabling the formation of high-density NV$^-$ ensembles without compromising structural integrity. Additionally, high-temperature implantation enhances NV$^-$ yield by up to five times compared to room-temperature implantation, while maintaining or even improving spin coherence properties. The yield is expected to be even higher for lower ion fluences.
These combined effects lead to a substantial enhancement of both AC and DC magnetic field sensitivity, making high-temperature implantation a promising strategy for quantum sensing applications.

We also demonstrated that even at 400°C, despite the amorphization threshold with fluence being identical to that at room temperature, a significant improvement in both the NV$^-$ creation yield and the NV$^-$ spin lifetime and coherence time is observed.

These results highlight the critical role of implantation temperature in optimizing  NV$^-$ ensembles for quantum sensing applications. 
Further insights in the NV$^-$ creation process could be gained by refining the implantation temperature in the range 300°C-600°C.
Future work could also focus on other parameters of the implantation process, including surface treatments to mitigate charge noise and improve spin coherence times \cite{sangtawesin_origins_2019}. 
Lower ion fluences should lead to a better yield and consequently better spin properties.
Additionally, expanding these studies to single NV$^-$ may provide deeper insights into the mechanisms governing NV$^-$s creation and coherence preservation \cite{pezzagna_creation_2010}.

\section*{Acknowledgments}
This work was funded by the European Research Council (ERC) under the European Union’s Horizon 2020 research and innovation program (RareDiamond, grant agreement No 101019234), by the regional network on quantum technologies in Ile-de-France (QuanTIP) and by the e-diamant (ANR Equipex+) project.

The authors would like to thank support from collaborators at Orsay Physics (Justine Renaud, Morgan Réveillard and Jérémie Silvent) as well as Ovidiu Brinza from LSPM-CNRS for diamond growth.

\bibliographystyle{elsarticle-num}
\bibliography{biblio.bib}

\appendix
\renewcommand{\thefigure}{\arabic{figure}}
\section{SRIM simulations}
\label{appendix:srim}
\begin{figure}[ht]
    \centering
    \includegraphics[width=0.7\columnwidth]{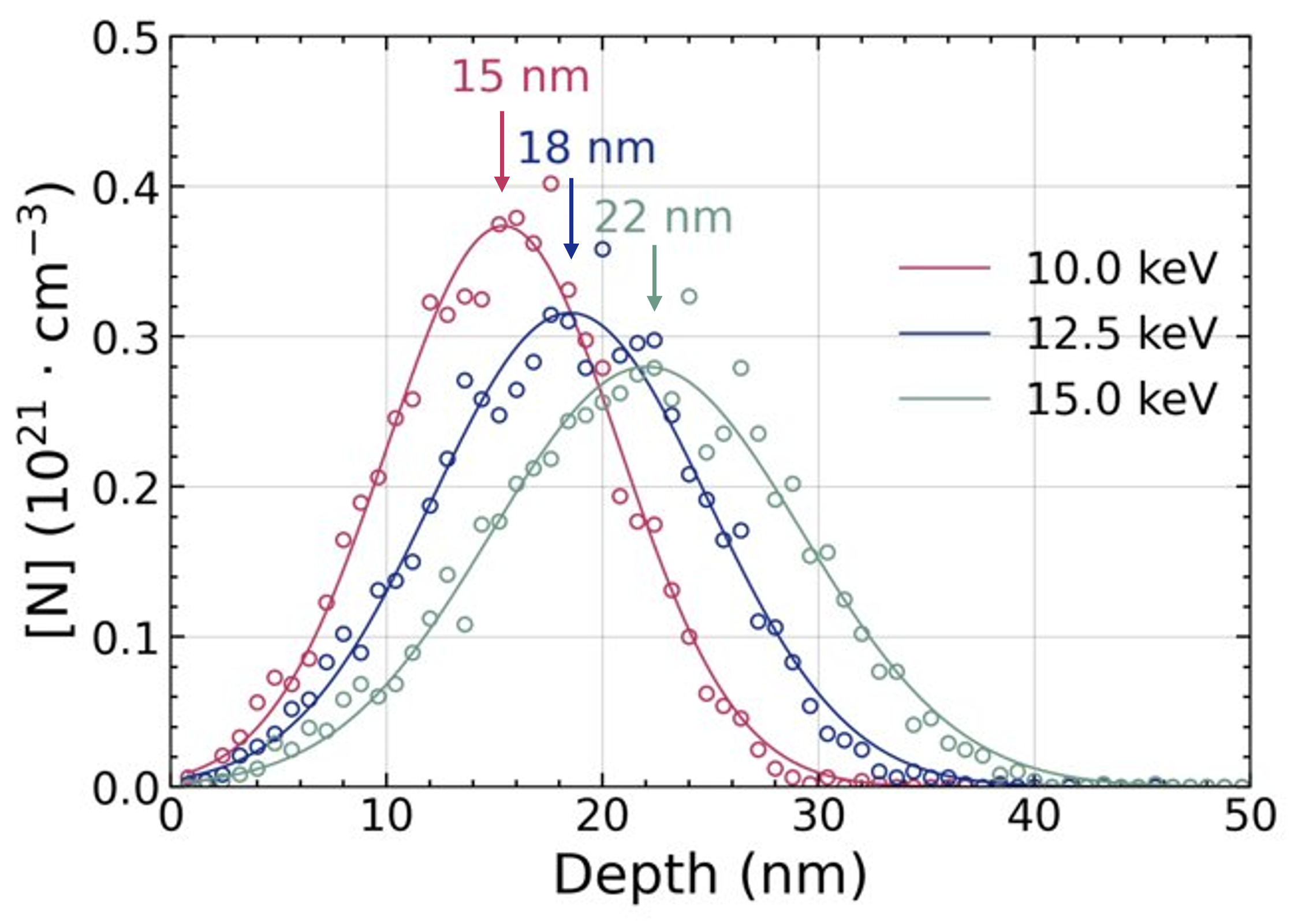}
    \caption{SRIM simulation of the depth distribution of N atoms for implantation energies of 10, 12.5, and 15 keV. The data are fitted using Gaussian functions.}
    \label{fig:srim}
\end{figure}

The depths of the NV$^-$ as a function of the implantation energy are simulated using the SRIM software \cite{ziegler_srim_2010}. In this paper, we study N$_2^+$ ions implanted at 20 keV, 25 keV and 30 keV. We consider that the N$_2^+$ dissociates in two N atoms of identical energies at the surface and have therefore  energies of 10, 12.5, and 15 keV.
The results are presented in Figure \ref{fig:srim}. The depths corresponding to the peak NV$^-$ concentrations are 15, 18, and 22 nm. Due to ion straggling, the NV$^-$ depth distribution is broadened as the energy is increased. The respective full widths at half maximum of these distributions are 12.6, 15.0 and 16.8 nm.

\section{Pulse sequences}
\label{appendix:pulselengths}

\begin{figure}[ht]
    \centering
    \includegraphics[width=\columnwidth]{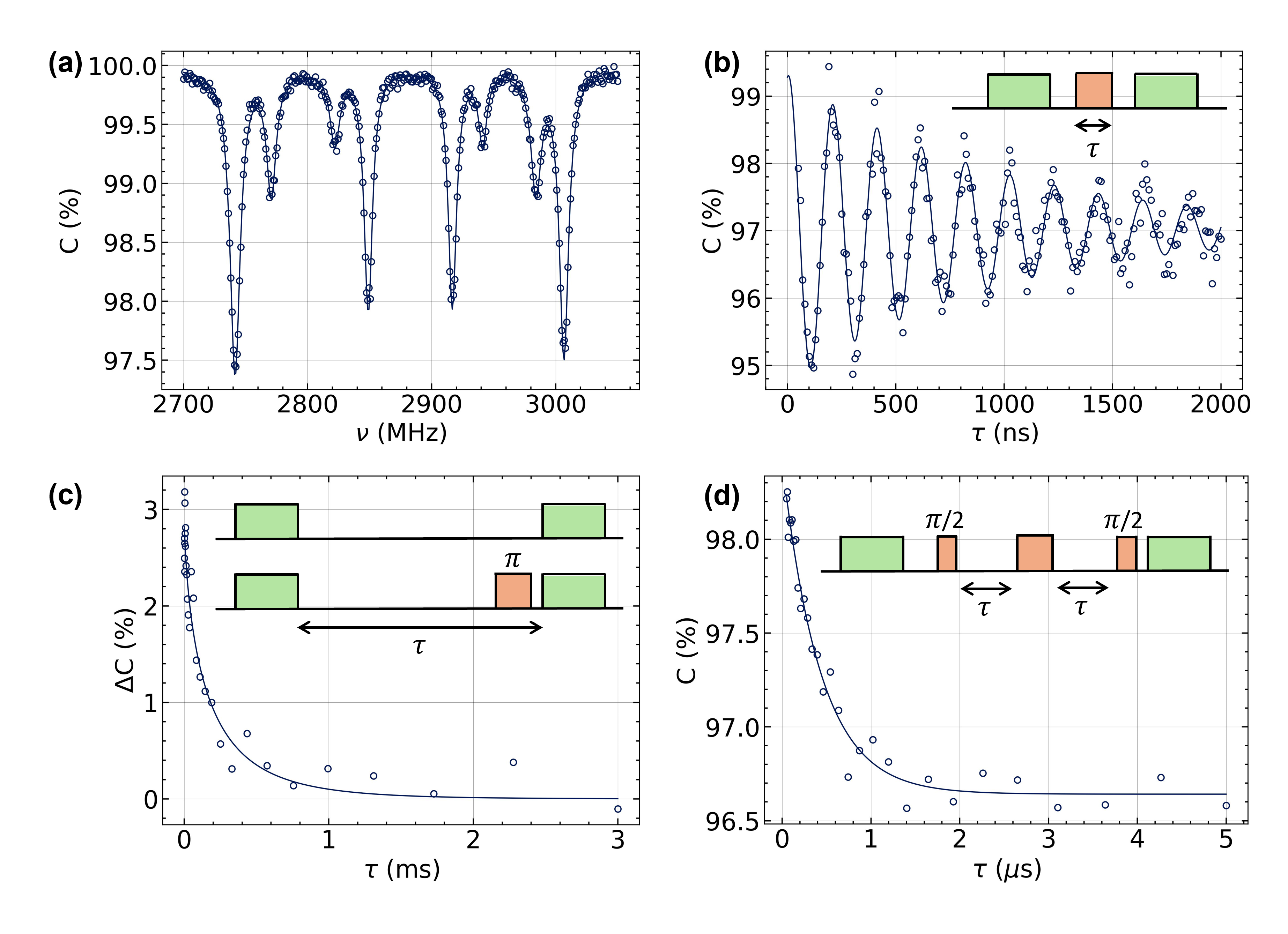}
    \caption{(a) ODMR spectrum for 30 dBm microwave power. The contrast is measured for various microwave frequencies. (b) Rabi oscillation. The contrast is measured for various microwave pulse durations $\tau$. (c) Lifetime measurement. The contrast is measured for various delays $\tau$ between the optical probe pulses. (d) Coherence time measurement. The contrast is measured for various delays $\tau$ between the $\pi$ pulse and the $\pi$/2 pulses.}
    \label{fig:fullmanip}
\end{figure}

To study the spin properties of the NV$^-$ ensemble, we began by recording the optically detected magnetic resonance (ODMR) spectrum. The PL of the NV$^-$ ensemble is recorded under continuous-wave laser excitation of 130 µW. To obtain the full ODMR spectrum, frequency of the applied microwave field is swept from 2700 to 3050 MHz.

The data is fitted using a sum of 8 Lorentzian functions.
To enhance the contrast of the targeted NV$^-$ family, we used linearly polarized optical excitation \cite{magaletti_magnetic_2024}. 
This method preferentially excites the NV$^-$ of interest while decreasing background fluorescence from other NV$^-$ sub-sites unaffected by the microwave excitation, thereby improving the signal-to-noise ratio.

For the Rabi frequency, T$_1$ and T$_2$ measurements, the NV$^-$ were initialized into the $\ket{m_s = 0}$ spin state via an optical pulse of duration $T_{\text{pump}} = 500~\mu$s. Following initialization, the NV$^-$ state was read out through fluorescence intensity using an optical pulse of duration $T_{\text{probe}} = 5~\mu$s. Afterward, the relevant microwave pulse sequence was applied, followed by a final optical readout pulse of duration $T_{\text{probe}} = 5~\mu$s to measure the final state of the NV$^-$ ensemble.
The durations of the $\pi$- and $\pi/2$-pulses, as well as the optimal values for $T_{\text{pump}}$ and $T_{\text{probe}}$, were chosen to maximize the amplitude of the Rabi oscillations.

To determine the optimal durations of the $\pi$ and $\pi$/2 pulses, we used the sequence shown in the inset of Figure \ref{fig:fullmanip}(b). Between the optical readout pulses, a microwave pulse of frequency $\nu$, resonant with the lowest-energy spin transition of the NV ensemble, was applied. The pulse duration $\tau$ was varied to investigate its effect on the luminescence contrast. The data, presented in Figure \ref{fig:fullmanip}(b), are fitted using an exponentially decaying sinusoidal function.

The pulse sequence employed to measure the ensemble spin lifetime is shown in the inset of Figure ~\ref{fig:fullmanip}(c). A delay $\tau$ was introduced between the first and second optical readout pulses. The sequence was performed twice: once with a $\pi$-pulse, yielding a measurement contrast $C_{\pi}$, and once without a $\pi$-pulse, yielding a measurement contrast $C_{\text{no }\pi}$.
The decay of the difference between these two contrasts, defined as \(\Delta C = C_{\text{no }\pi} - C_{\pi}\), isolates the decay of the NV$^-$ spin state of interest. The resulting decay curves were fitted with a $\beta$ stretched exponential function.
The inclusion of the stretching parameter $\beta$ accounts for the distribution of NV$^-$ depths and the presence of cross-relaxation effects within the NV$^-$ ensemble.

The pulse sequence employed to measure the ensemble spin coherence time is shown in the inset of Fig.~\ref{fig:fullmanip}(d). 
The sequence begins with a microwave $\pi/2$-pulse, which creates a coherent superposition state. 
After a time delay $\tau$, a $\pi$-pulse is applied to refocus the spin ensemble.
The system then evolves for an additional delay $\tau$, after which a second $\pi/2$-pulse converts the refocused coherence into a population signal.
A typical measurement is presented in Fig.~\ref{fig:fullmanip}(d). 
The luminescence contrast is recorded for various delays $\tau$, and the Hahn echo decay is fitted using a mono-exponential function.
\section{\texorpdfstring{Reference sample for NV$^{-}$ concentration}{Reference sample for NV- concentration}}
\label{Appendix E}

To estimate the NV$^-$ creation yield for the different implantation conditions, we used a reference sample with known NV$^-$ concentrations.
It consists of a 4-layer CVD diamond film grown onto an HPHT substrate \cite{balasubramanianEnhancementCreationYield2022}. While the first layer A contains no intentional doping, the layers B, C and D are grown with the addition of respectively 10, 20 and 50 ppm of N$_2$O in the growth chamber.

The NV$^-$  concentration in each layer was calculated by comparing the fluorescence intensity in each layer to the fluorescence intensity of a single NV$^-$. The resulting NV center concentrations of layers B, C and D are of 358, 530 and 835 ppb respectively.


\end{document}